\documentclass[a4paper,10pt,twoside]{cpc-hepnp}

\usepackage{multicol}
\usepackage{graphicx}
\usepackage{booktabs}
\usepackage{amssymb,bm,mathrsfs,bbm,amscd}
\usepackage[tbtags]{amsmath}
\usepackage{lastpage}

\begin{document}

\fancyhead[c]{\small Submitted to 'Chinese Physics C'
} \fancyfoot[C]{\small 010201-\thepage}

\footnotetext[0]{Received XX June 2014}

\title{Photon activation analysis of the scraper in a 200-MeV electron accelerator using gamma-spectrometry depth profiling\thanks{Supported by National Natural Science Foundation of China (11175180 and 11175182) }}

\author{%
      He Lijuan(ºÎÀö¾ê)$^{1;1)}$\email{juan@mail.ustc.edu.cn}%
\quad Li Yuxiong(ÀîÔ£ÐÜ)$^{1;2)}$\email{lyx@ustc.edu.cn}%
\quad Yu Guobing(춹ú±ø)$^{2}$
\quad Ren Guangyi(ÈιãÒæ)$^{1}$
\quad Duan Zongjin(¶Î×Ú½õ)$^{1}$
}
\maketitle

\address{%
$^1$ National Synchrotron Radiation Laboratory, University of Science and Technology of China, Hefei, Anhui 230029, P.R.China\\
$^2$ Radiation Monitoring Station, Hefei, Anhui, 230071, China\\
}

\begin{abstract}
For a high energy electron facility, the estimates of induced radioactivity in materials are of major importance to keep exposure to personnel and to the environment as low as reasonably achievable. In addition, an accurate prediction of induced radioactivity is also essential for the design, operation and decommissioning of a high energy electron linear accelerator. The research of induced radioactivity focuses on the photonuclear reaction, whose giant resonance response in the copper is ranging from 10 MeV to 28 MeV. The 200 MeV electron linac of NSRL is one of the earliest high-energy electron linear accelerators in P. R. China. The electrons are accelerated to 200 MeV by five acceleration tubes and collimated by the scrapers made of copper. The energy of five acceleration tubes is 32, 74, 116, 158 and 200 MeV respectively. At present, it is the first retired high-energy electron linear accelerator in domestic. Its decommissioning provides an efficient way for the induced radioactivity research of such accelerators, and is a matter of great significance to the accumulation of the induced radioactivity experience. When the copper target is impacted by an 158 MeV electron beam, the number of photons generated whose energy are in the range of giant resonance response is the largest. Thus, this paper focuses on the induced radioactivity for a copper target impacted by the 158 MeV electron beam. The slicing method is applied in the research. The specific activity of each slice was measured at cooling times of ten months and the results were compared with the prediction from the Monte-Carlo program FLUKA. The simulation results are in good agreement with the measurement results. The method by Monte Carlo simulation in this paper gives a reasonable prediction of the induced radioactivity problem for the high-energy electron linear accelerators, laying a foundation for the accumulation of the induced radioactivity experience.
\end{abstract}

\begin{keyword}
NSRL Linac, decommissioning, induced radioactivity, scraper, 158 MeV electron beam, $^{60}$Co, HPGe ¦Ã spectrometer, FLUKA
\end{keyword}

\begin{pacs}
29.20.Ej, 24.10.Lx, 25.20.-x
\end{pacs}

\footnotetext[0]{\hspace*{-3mm}\raisebox{0.3ex}{$\scriptstyle\copyright$}2013
Chinese Physical Society and the Institute of High Energy Physics
of the Chinese Academy of Sciences and the Institute
of Modern Physics of the Chinese Academy of Sciences and IOP Publishing Ltd}%

\begin{multicols}{2}

\section{Introduction}

The Synchrotron Radiation Facility of National Synchrotron Radiation Laboratory (NRSL), located in Hefei, China, is one of the earliest synchrotron light sources in China. NRSL is the home to a 200-MeV electron linear accelerator with a length of 35 m as the injector and an 800-MeV electronic storage ring\cite{lab1}. Constructed in 1989, the 200-MeV electron linac is a traveling wave linear accelerator, consisting of five acceleration tubes (in which the electrons are accelerated to 32, 74, 116, 158 and 200 MeV, respectively).

The NSRL Linac was retired in 2012 due to a plan to upgrade. The induced radioactivity along with its decommissioning is one of the major concerns\cite{lab2}. When accelerated electrons are lost during machine operations, such high-energy electrons will cause electromagnetic cascades while colliding with materials. The resulting high-energy bremsstrahlung will induce photonuclear reactions. Photon activation usually dominates the activation process that occurs in electron accelerators. Furthermore, secondary neutrons produced through photo-nuclear reactions may contribute to the activation to a certain extent.

Induced radioactivity for other types of accelerators has been investigated and was found to contain high levels. Wu et al. (2011)\cite{lab3} discussed the induced radioactivity of China Spallation Neutron Source. Adopting the FLUKA code, the authors calculated air activation and analyzed various issues with regard to the activation of different tunnel parts. The results show that the residual radiation in the tunnels has a great influence on maintenance personnel. Bi et al. (2009)\cite{lab4} studied the prompt radiation and residual radiation field for CYCIAE-100 using Monte Carlo method. Meanwhile, the radioactive contamination near stripping foil was researched and a method to reduce the dose equivalent rate of maintenance staff was also given. Wang et al. (2007)\cite{lab5} evaluated the residual radioactivity induced in shielding concrete of accelerator facilities using Monte Carlo method EGS4 code and NaI survey meter. Li et al. (1991)\cite{lab6} analyzed and discussed the data and spectrum acquired from the induced radiation field in NSRL 200 MeV linac tunnel.

Residual radioactivity can lead to three radiation safety concerns: personnel exposure, environmental impact, and waste disposal\cite{lab7}. In a well-shielded accelerator facility, a great portion of the personnel dose arises from exposure to radiation of activated components during maintenance work. Meanwhile, possible environmental release of gaseous and liquid radionuclides is a sensitive issue. Furthermore, the disposal of activated materials and the decommissioning of accelerators need to be carefully evaluated.

Induced radioactivity may be created in various accelerator components and accelerator's surroundings when irradiated directly by the primary beam or secondary radiation fields. For the NSRL Linac, the scraper, which is made of copper and used for beam collimation, is located after each acceleration tube, and the length of the scraper increases with the energy of the electrons. The scraper aperture has a diameter of 1cm, which is smaller than that of the acceleration tube which has a diameter of\cite{lab8}. Therefore, it is reasonable that the electrons mainly loss at the scraper due to the energy dispersion when being transported and accelerated. Based on this reason, the induced radioactivity is mainly produced at the scraper, and the handling of the structural materials after the linac has been retired will become easier. That is to say the handling of the scrapers is the focus of the decommissioning. Furthermore, this hypothesis has been proved by the measurements. The accelerator structures besides the scrapers have been measured, and the measurements showed that the residual radioactivity in the structures is negligible compared with the scrapers.

All the radioactivity of the scraper in this study originates from electromagnetic cascades and subsequent nuclear reactions. For a 200 MeV electron linear accelerator, the main nuclear reactions are introduced by the photons, such as (¦Ã,n), (¦Ã,2n), (¦Ã,np), (¦Ã,n2p) and so on\cite{lab9}. The cross section of (¦Ã,n) is the biggest, and the ones of (¦Ã,2n) and (¦Ã,np) are followed, finally (¦Ã,n2p) has the least cross section. For the copper materials, the giant resonance region of the photonuclear reaction is about 10-28 MeV.

\section{Materials and methods}

\subsection{Introduction to the scraper}

For the NSRL Linac, the electrons are accelerated up to the expected energy in each acceleration tube and collimated by the scraper followed each acceleration tube. The radius of the scraper is 4.8cm, and the pore diameter is 1cm. While the pore size of disk-loaded waveguide is 2.1977cm. The scraper aperture is smaller than the acceleration tube one, so some electrons will impact on the structure materials due to the energy dispersion when they pass through them.

The electrons are accelerated to 158 MeV in the fourth acceleration tube, and are then collimated in the fourth scraper. The fourth scraper is an annular copper cylinder (9.6cm diameter¡Á9.0cm height¡Á3.8cm wall thickness). The actual structure is shown in Fig.~\ref{fig1}.
\begin{center}
\includegraphics[width=6.8cm]{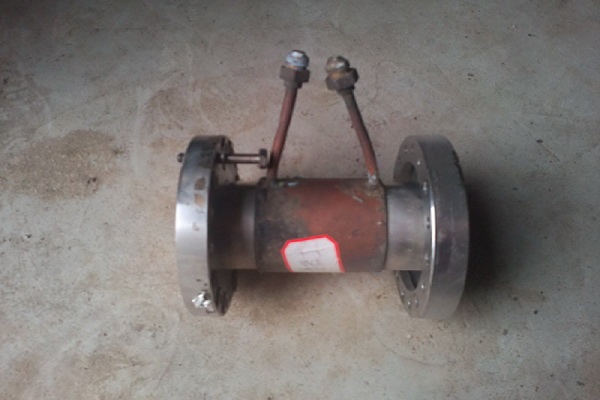}
\figcaption{\label{fig1}   Structure of the fourth scraper (T2 copper). }
\end{center}

\subsection{Measurement of radionuclides}

To study the induced radioactivity of the fourth scraper, its ¦Ã energy spectrum has been measured many times when the accelerator was shut down. The measuring system is a portable HPGe ¦Ã spectrometer (GR3519 with a relative efficiency of 35\% and an energy resolution of 1.9 keV).

\subsection{Specific activity measurement of $^{60}$Co in sliced coppers}

The radionuclide in the fourth scraper mostly is $^{60}$Co after the machine shutdown for about ten months. To acquire a formula that describes the relationship between the depth and the specific activity of $^{60}$Co, the slicing method is applied in the research. The actual number of slice is at the bottom of the scraper. The thicknesses of the first and eighth sheet are both 1.5cm, and the others are 1cm. The photo of the sliced copper is shown in Fig.~\ref{fig2}. The ¦Ã energy spectra of eight sliced coppers have been measured by a P-type HPGe ¦Ã spectrometer (GC4019 with a relative efficiency of 40\% and an energy resolution of 1.9 keV).
\begin{center}
\includegraphics[width=6.8cm]{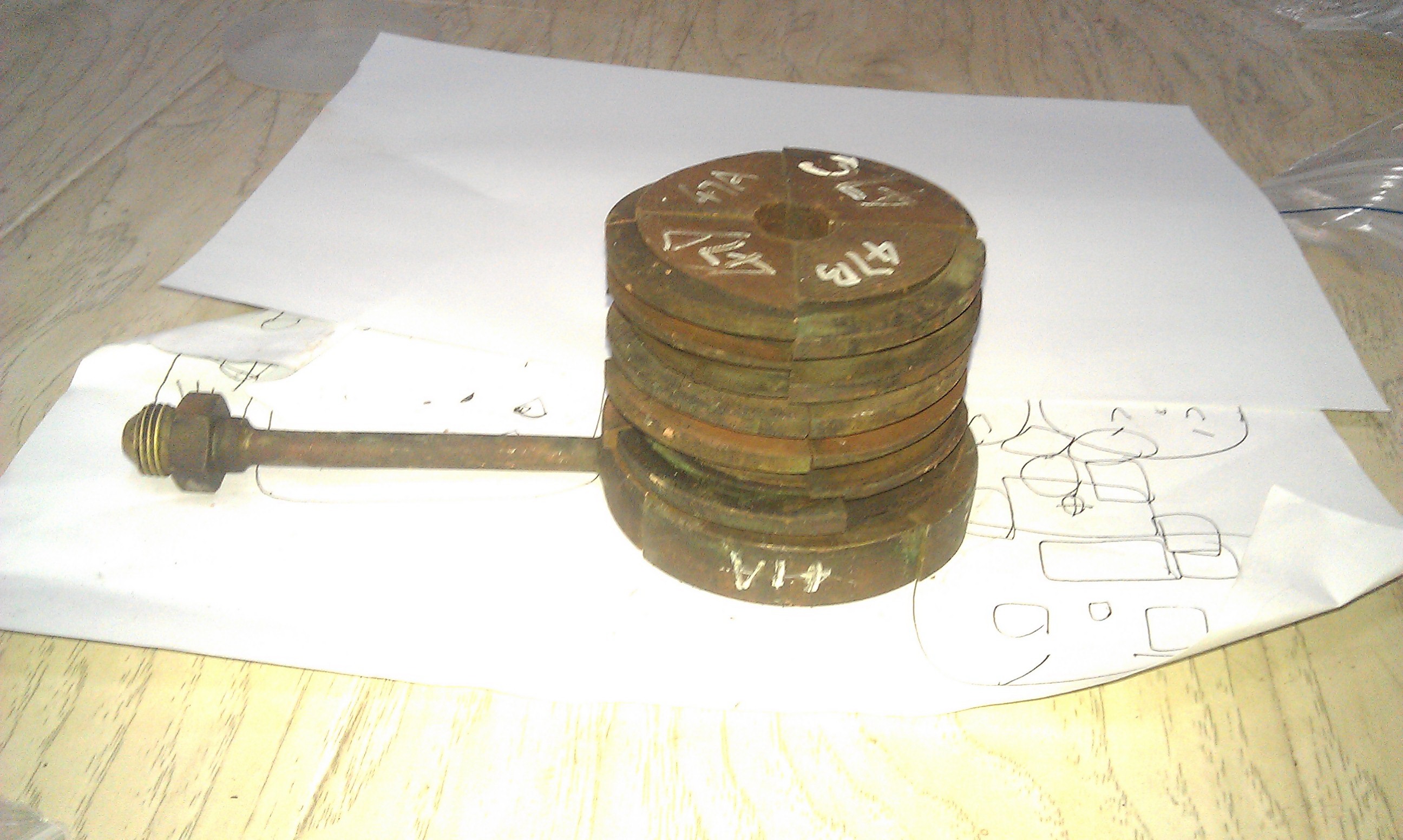}
\figcaption{\label{fig2}   Photo of sliced coppers. }
\end{center}

\subsection{Simulation of the yield of photon and $^{60}$Co in sliced coppers}

The research has shown that the photonuclear reaction which produces $^{60}$Co is $^{63}$Cu(¦Ã,n2p)$^{60}$Co, and the threshold of the reaction is 18.86 MeV\cite{lab5}. The energy range of its giant resonance response hasn't been found in the existing literature; hence the energy upper limit (about 28 MeV) of giant resonance photonuclear reaction in the copper\cite{lab5} is regarded as its upper threshold. The photon yield within energy from 18.86 to 28 MeV in average volume of each sliced copper is recorded, when the electron beam with 158 MeV impacts on a copper target in the simulation. Furthermore, the yield of $^{60}$Co is simulated by Monte Carlo program FLUKA\cite{lab10}. FLUKA is now capable of giving estimates of the number and types of radionuclides created by photonuclear reactions at high-energy electron linear accelerator.

The initial conditions for the simulations are as follows. The electron energy is 158 MeV, and the beam is annular (1.6cm diameter¡Á0.3cm wall thickness). The copper target is an annular cylinder (9.6cm diameter¡Á9.0cm height¡Á3.8cm wall thickness). The photon yield within threshold and the yield of $^{60}$Co in eight sliced coppers are recorded. The simplified model is shown in Fig.~\ref{fig3}.
\begin{center}
\includegraphics[width=6.8cm]{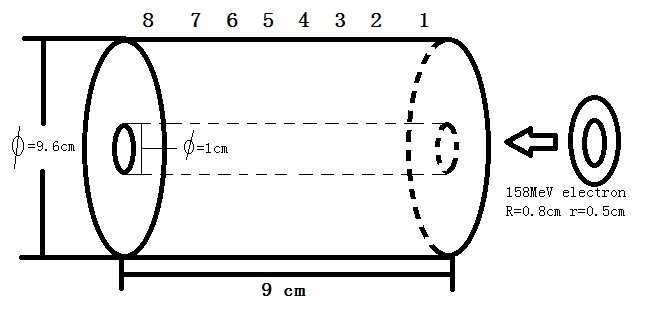}
\figcaption{\label{fig3}   Simplified model of the fourth scraper. }
\end{center}

\section{Results and discussion}

The measured radionuclides are: $^{57}$Ni, $^{52}$Mn, $^{51}$Cr, $^{58}$Co, $^{56}$Co, $^{57}$Co, $^{54}$Mn, $^{22}$Na and $^{60}$Co. The details related to the radionuclides are seen in another article ¡°Induced radioactivity research for scraper¡±\cite{lab11}.

The specific activity of $^{60}$Co in eight sliced coppers is shown in Fig.~\ref{fig4}. The specific activity of $^{60}$Co in the second is the greatest. By the research and measurement, one can see that the radionuclide emitting ¦Ã rays with the longest half-life is $^{60}$Co in the scraper. In order to study the relationship between the yield of $^{60}$Co and the depth, related simulations are performed. The simulation result of photon yield is shown in Fig.~\ref{fig5}. The cross section data of the reaction has not been found in the related literature and web site. Yet according to the related cross section data of the photonuclear reaction in copper, it can be inferred that its cross section data is quite small. For the same material and reaction, the content of $^{63}$Cu and the reaction cross section data are the same, so the yield of $^{60}$Co can be estimated according to the photon yield within the threshold. The simulation result indicates that the photon yield in the second piece is the most, which can reflect that the yield of $^{60}$Co is the most in the second to a certain extent. Furthermore, the simulation of the $^{60}$Co yield in eight sliced coppers is performed, and the result is shown in Fig.~\ref{fig6}. The prediction shows that the data in the second sliced copper is the highest, which agrees well with Fig.~\ref{fig5}. The result indicates that the estimated method of $^{60}$Co yield according to the photon yield within threshold is feasible and very efficient.
\begin{center}
\includegraphics[width=6.8cm]{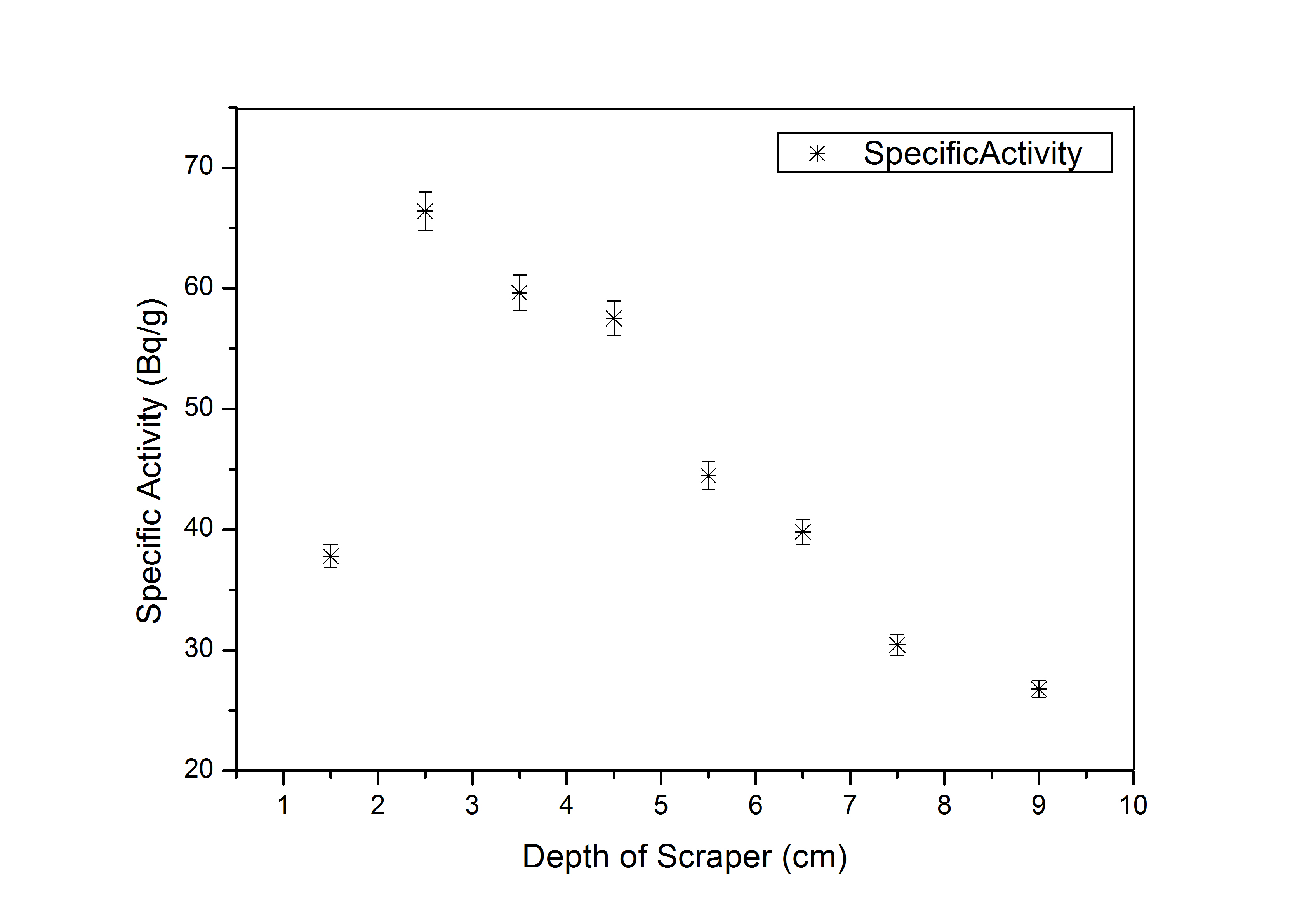}
\figcaption{\label{fig4}   Specific activity of $^{60}$Co in each sliced copper. }
\end{center}
\begin{center}
\includegraphics[width=6.8cm]{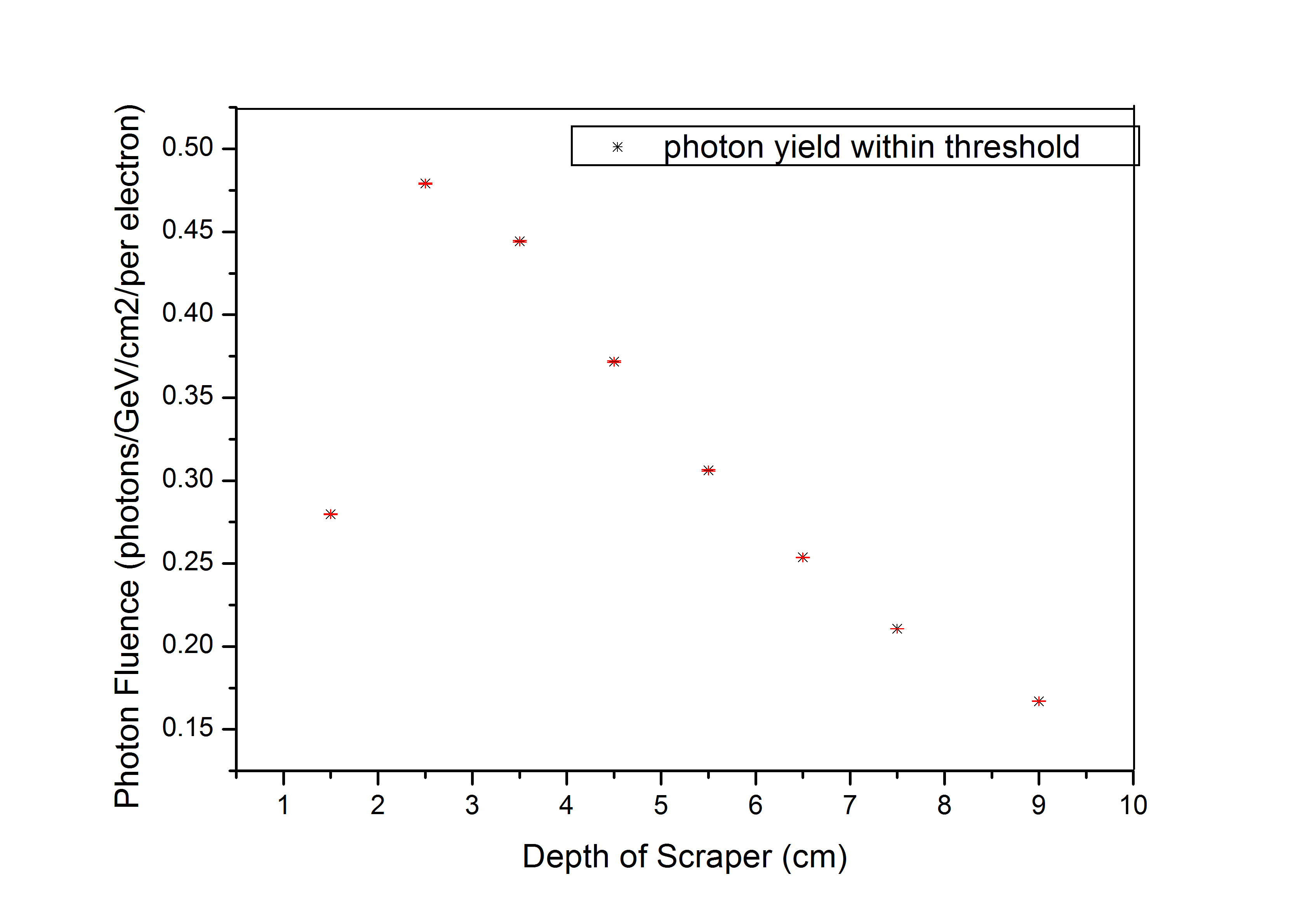}
\figcaption{\label{fig5}   Photon yield within threshold in each piece. }
\end{center}
\begin{center}
\includegraphics[width=6.8cm]{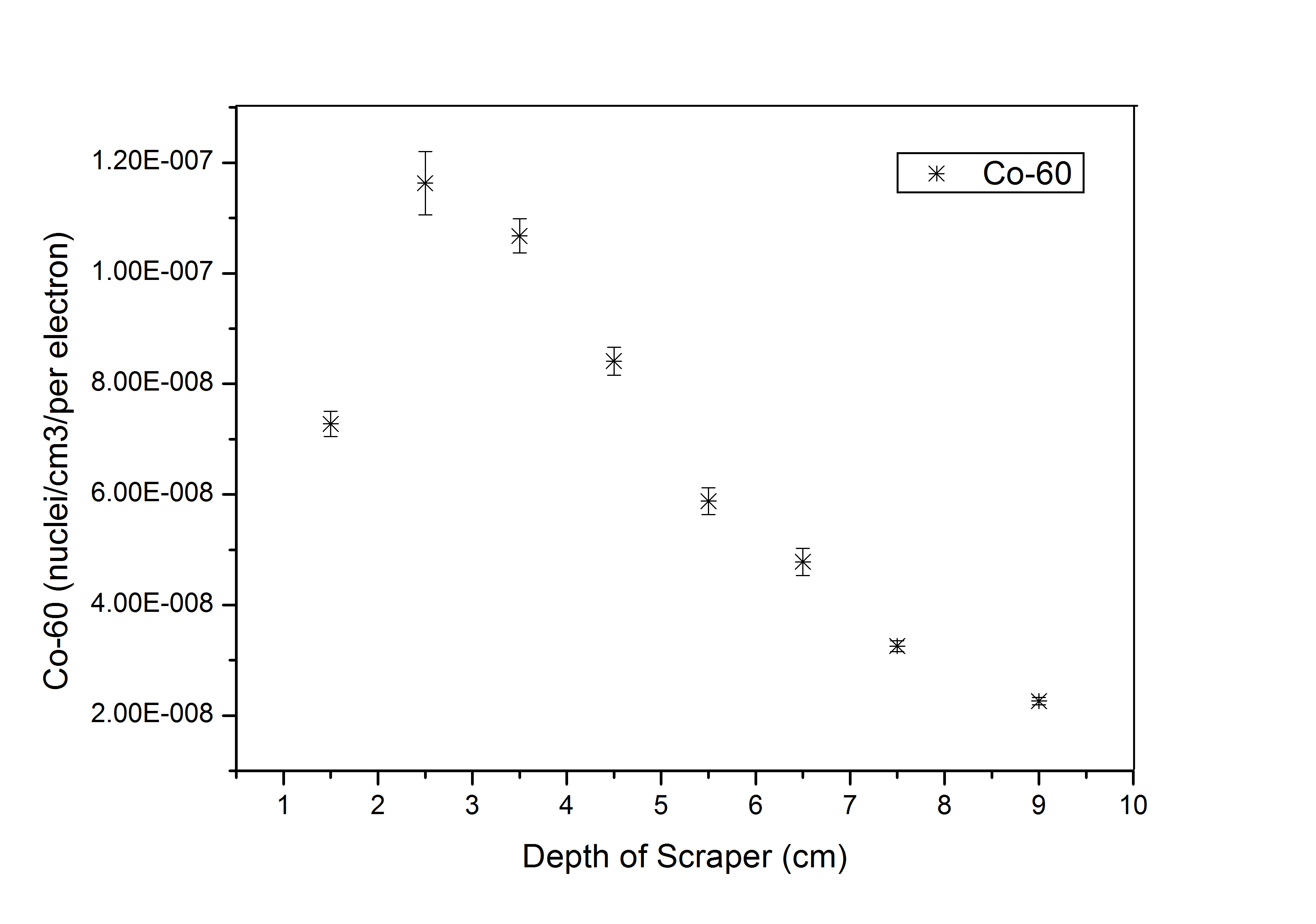}
\figcaption{\label{fig6}    Yield of $^{60}$Co in each sliced copper. }
\end{center}

It can be seen that the number of photons in the second sliced copper within threshold is the most, showing good agreement with the measurement result. The factors affecting the $^{60}$Co yield are: the content of $^{63}$Cu, the cross section and the photon yield within the threshold. For the same material, the content of $^{63}$Cu is the same; and for the same reaction, the data of cross section is the same. Based on the two reasons, the amount of $^{60}$Co generated mainly depends on the photon yield within threshold. Therefore, the data in Fig.~\ref{fig5} can reflect the yield of $^{60}$Co in each sliced coppers to a certain extent. Meanwhile, the $^{60}$Co yield is predicted using the FLUKA simulation. The simulation results about the relationship between the yield of $^{60}$Co and the depth are consistent with the measurement results.

\section{Conclusions}

For an electron linear accelerator, the scraper is mainly used for collimating the beam from the physical viewpoint. Considering the radiation protection point of view, the decommissioning of the machine and the dismantling of the structural material become feasible and convenient because the presence of the scraper.

In this work, radionuclides generated in a copper target (impacted by a 158 MeV electron beam) mainly conclude $^{57}$Ni, $^{52}$Mn, $^{51}$Cr, $^{58}$Co, $^{56}$Co, $^{57}$Co, $^{54}$Mn, $^{22}$Na and $^{60}$Co, etc. It has shown that the radionuclide emitting ¦Ã rays with the longest half-life is $^{60}$Co. The photonuclear reaction generated $^{60}$Co is $^{63}$Cu (¦Ã,n2p)$^{60}$Co, the threshold of which is 18.86-28 MeV. The slicing method is applied to obtain the depth profiling of $^{60}$Co in the scraper. The measurement indicates that the specific activity of $^{60}$Co in the second sliced copper is the most.

In order to obtain the yield trend of $^{60}$Co with the depth, some simulations have been performed using FLUKA. It can be found that the photon yield within threshold in the second has the maximum value in the simulation. Besides, the simulation result shows that the yield of $^{60}$Co in the second sliced copper is the most. The influential factors of the $^{60}$Co yield include the content of $^{63}$Cu, the cross section and the photon yield. The yield of $^{60}$Co only depends on the photon yield within threshold when the first two factors are the same. That is to say, the $^{60}$Co emission can be predicted by simulating the photon yield. Moreover, the simulation results about the relationship between the yield of $^{60}$Co and the depth are consistent with the measurements.

Monte Carlo simulation in this paper gives a reasonable prediction of the $^{60}$Co yield trend with the depth in the scraper. Furthermore, the slicing method is useful and efficient in the depth profiling of radionuclide, laying a foundation for the accumulation of the induced radioactivity experience in the electron accelerators.
\\

\acknowledgments{The authors would like to thank X. George Xu for his assistance in the process of article revision.}

\end{multicols}

\vspace{-1mm}
\centerline{\rule{80mm}{0.1pt}}
\vspace{2mm}

\begin{multicols}{2}

\end{multicols}

\clearpage

\end{document}